\documentclass[reprint,showkeys,nofootinbib,amsmath,amssymb,aps,prd,]{revtex4-1}
\usepackage[marginal]{footmisc}
\usepackage{graphicx}
\usepackage{dcolumn}
\usepackage{bm}
\usepackage[colorlinks,allcolors=blue,CJKbookmarks=True]{hyperref}
\usepackage{amsmath}
\usepackage{amssymb}
\usepackage{subfigure}
\usepackage{amsfonts}
\usepackage{physics}
\usepackage{upgreek}
\usepackage{amssymb}
\usepackage{dcolumn}
\usepackage{bm}
\usepackage{tikz}
\usepackage{graphicx}
\usepackage{xcolor}
\usepackage{array}
\usepackage{verbatim}
\usepackage{hyperref}
\newcommand{\dif}{\mathrm{d}}
\newcommand\mi{\mathrm{i}}
\newcommand\me{\mathrm{e}}

\newcommand\pp{\uppi}

\begin{document}
	\title{Spectrum of third-order tensor perturbations induced by excited scalar fields}

\author{Cheng-Jun Fang$^{1,2}$}
\email{fangchengjun@itp.ac.cn}
	
\author{Zhi-Zhang Peng$^{4}$}
\email{pengzhizhang@bnu.edu.cn}
	
\author{Han-Wen Hu$^{1,2}$}
\email{huhanwen@itp.ac.cn}	
	
\author{Zong-Kuan Guo$^{1,2,3}$}
\email{guozk@itp.ac.cn}
	
	\affiliation{$^1$CAS Key Laboratory of Theoretical Physics, Institute of Theoretical Physics, Chinese Academy of Sciences, P.O. Box 2735, Beijing 100190, China}
	
	\affiliation{$^2$School of Physical Sciences, University of Chinese Academy of Sciences, No.19A Yuquan Road, Beijing 100049, China}

	\affiliation{$^3$School of Fundamental Physics and Mathematical Sciences, Hangzhou Institute for Advanced Study, University of Chinese Academy of Sciences, Hangzhou 310024, China}
	\affiliation{$^4$School of Physics and Astronomy, Beijing Normal University, Beijing 100875, People's Republic of China}
\begin{abstract}
We calculate for the first time the third-order spectrum of gravitational waves sourced by the scalar field perturbations amplified during inflation using the in-in formalism, and discuss the conditions for the third-order spectrum to be smaller than the second-order one. We find that the third-order spectrum increases faster than the second-order one as the amplification of the field perturbations increases, and thus the third-order spectrum dominates for detectable gravitational waves, which indicates that the perturbation theory breakdowns.
\end{abstract}

\maketitle
\emph{Introduction}.
The advanced LIGO detectors opened the era of gravitational-wave (GW) astronomy with the first detection of GWs
from a merger of binary black holes~\cite{LIGOScientific:2016aoc}.
GW observations allow us to understand the physical processes happened in the early Universe, 
especially inflation~\cite{Boyle:2005se,Guzzetti:2016mkm,Bartolo:2016ami}. 
Some inflationary mechanisms,
such as ultra-slow-roll inflation~\cite{Namjoo:2012aa,Garcia-Bellido:2017mdw,Germani:2017bcs,Motohashi:2017kbs,Xu:2019bdp,Fu:2019vqc} and sound speed resonance~\cite{Cai:2018tuh,Cai:2019jah,Cai:2019bmk},
have been proposed to generate strong GW signals, which are expected to be detected by future laser interferometers.

This Letter focuses on an inflationary scenario in which the scalar field perturbations are enhanced during inflation. 
A typical mechanism for generating these amplified scalar modes is parametric resonance~\cite{Cai:2018tuh,Cai:2019jah,Cai:2019bmk,Zhou:2020kkf,Cai:2021wzd,Inomata:2022yte}.
Those excited scalar modes can source second-order GWs during inflation ~\cite{Fumagalli:2021mpc, Peng:2021zon}. 
Meanwhile, small-scale curvature perturbations are also enhanced,
which lead to the production of second-order GWs when entering the horizon in the radiation-dominated (RD) era~\cite{Baumann:2007zm, Domenech:2021ztg, Fumagalli:2020nvq}. 
Thus, in the scenario there are two peaks in the energy spectrum of the stochastic GW background~\cite{Peng:2021zon}.

Following standard cosmological perturbation theory, we can expand GW as $h_{ij}=h^{(1)}_{ij}+\frac{1}{2}h^{(2)}_{ij}+\frac{1}{6}h^{(3)}_{ij}+\cdot\cdot\cdot$,  where $h^{(n)}_{ij}$ is the $n$-th order GW and we temporarily omit the spin index. Traditionally, $\left \langle h^{(2)}_{ij}h^{(2)}_{ij}\right \rangle$ are always considered as a good approximation of the physical results. 
For GWs induced during the RD era, higher order contributions are small, but it’s still observable \cite{Yuan:2019udt,Chang:2022nzu,Chang:2023vjk} in come conditions. For example, the cross-correlation $\left \langle h^{(1)}_{ij}h^{(3)}_{ij}\right \rangle$ can slightly reduce the GWs amplitude at large scales \cite{Chen:2022dah}. Recently,  one-loop tensor power spectrum generated during inflation is also fully considered \cite{Ota:2022hvh,Ota:2022xni}, where the cross-correlations gives a large contribution to the power spectrum, especially at the infrared limit. However, there remains a challenge to the validity of perturbative calculations, just as in the case of ultra slow roll\cite{Kristiano:2022maq,Kristiano:2023scm} and parameter resonance\cite{Inomata:2022yte}. For tensor modes, whether these one-loop order spectrum can be considered as physical approximations remains an unresolved issue.

In this Letter, we  calculate for the first time the third order corrections to power spectrum of GWs in the case of large scalar field perturbations during inflation. Corresponding to the order expansion of tensor perturbations, we calculate the power spectrum of third-order GWs, i.e., $\left \langle h^{(3)}_{ij}h^{(3)}_{ij}\right \rangle$.  We focus solely on the infrared corrections because the infrared behavior of GW power spectrum exhibits unique scale-invariant corrections in the one-loop contributions and is also constrained by large-scale observations. We find that the growth of the third order spectrum exhibits a greater dependence on the amplification parameter than that of the second order ones and finally dominate over one-loop contributions which indicates the perturbative method has broken down. Surprisingly, we discover that the GW power spectrum corresponding to the critical magnification factor remains undetectable by GW detectors such as LISA, Taiji, and TianQin.

Moreover, our analysis is model-independent and, at the perturbative level, provides stronger constraints on model parameters compared to scalar field perturbations. We have discussed the conditions under which GWs can be generated perturbatively. Constraints given by these conditions are extremely strict, almost forbid the perturbative production of detectable GWs during inflation.

\emph{Set-up}.
There are many modes can generate larger scalar perturbations, which subsequently sources GWs. However, to provide a model-independent analysis, this paper will not focus on specific models. Instead, we will consider a spectator field $\delta\phi$ minimally coupled to gravity and parameterize the enhancement of its mode functions. We can write down corresponding part of the action  as 
\begin{equation}
	S_{\delta\phi}=\frac{M_{\rm p}^2}{2}\int \dif ^4 x \sqrt{-g}\mathcal{R}+\int \dif ^4 x \sqrt{-g}\mathcal{L}_{\delta\phi}.
\end{equation}
To perturbatively expand this action to higher orders, ADM decomposition of the metric is commonly employed
\begin{equation}
	\dif s^2=-N^2 \dif \tau^2+\gamma_{i j}\left(\dif x^i+N^i \dif \tau\right)\left(\dif x^j+N^j \dif \tau\right),
\end{equation}
where $N$ and $N_i$ are the lapse function and shift vector, respectively, serving as Lagrange multipliers, $\tau$ is the conformal time, and $\gamma_{ij}$ is the spatial components of the metric. Since the mapping from background spacetime to perturbed spacetime is not unique, there are gauge freedoms need to be fixed. In our calculations we choose the uniform curvature gauge where the field perturbations $\delta\phi$ act as the dynamical variable. Following Maldacena\cite{Maldacena:2002vr}, the spatial parts are written as 
\begin{equation}
	\gamma_{i j}=a^2 \me^{h_{i j}},
\end{equation}
where $h_{i j}$ are tensor perturbations which are transverse $\left(\partial^i h_{i j}=0\right)$ and traceless $\left(\delta^{i j} h_{i j}=0\right)$. 
We expand the action to fourth order and solve the constrains to obtain $N$ and $N_i$, which are both suppressed by slow-roll parameters in spatially flat gauge\cite{Inomata:2022yte,Inomata:2024lud}. As a result, these metric perturbations can be safely neglected. With our definition of $h_{i j}$, we have $\operatorname{det}\left|e^{h_{i j}}\right|=1$, which leads to the fact that tensor perturbations keeps the volume element at non-linear order. Therefore, only kinetic terms of the field Lagrangian contributes to the interaction with tensor parts. Applying Legendre transformation to the action, we arrive at the interaction Hamiltonian
	\begin{align}
		H_{\text {int }} &\equiv H^{(3)}+H^{(4)} \nonumber\\
		&=a^2 \int \dif^3 x\left(-\frac{1}{2} h^{i j}+\frac{1}{4} h^{i k} h_k{ }^j\right) \partial_i \delta\phi \partial_j \delta\phi.
	\end{align}
Our goal is calculating the primordial tensor power spectrum
\begin{align}
	& \left.\left\langle \Omega\left|\sum_s h_{\mathbf{q},\rm H}^s(\tau) h_{\mathbf{q}^\prime,\rm H}^s(\tau)\right| \Omega\right\rangle\right|_{\tau=0} \nonumber \\
    &\equiv(2 \pp)^3 \delta(\mathbf{q}+{\mathbf{q}^\prime}) P,
\end{align}
where $h_{\mathbf{q},\rm H}^s(\tau)$ is a operator in the Heisenberg picture. Vacuum expectation value of the  operator $\mathcal{O}$ can be computed with in-in formalism via the following formula
\begin{equation}
	\langle\mathcal{O}\rangle= \lim_{\tau_i \rightarrow-\infty(1- \mi \epsilon)}\langle 0| F(\tau ; \tau_i)^\dagger 
	\mathcal{O}_{\rm I}(\tau)F\left(\tau ; \tau_i\right) |0\rangle,
\end{equation}
where $F(\tau;\tau_0)$ denotes the evolution operator in the interaction picture, which can be solved iteratively and the Heisenberg picture operator can then be written as
	\begin{align}
		&h_{\mathbf{q},\rm H}^s(\tau)=F\left(\tau ; \tau_i\right)^{\dagger} h_{\mathbf{q}}^s F\left(\tau ; \tau_i\right) \nonumber\\
		&=h_{\mathbf{q}}^s(\tau)+\mi \int_{\tau_i}^\tau \dif \tau^{\prime}\left[H_{\mathrm{int}, {\rm I}}\left(\tau^{\prime}\right), h_{\mathbf{q}}^s(\tau)\right] \nonumber\\
		& -\int_{\tau_i}^\tau \dif \tau^{\prime} \int_{\tau_i}^{\tau^{\prime}} \dif \tau^{\prime \prime}\left[H_{\mathrm{int}, {\rm I}}\left(\tau^{\prime \prime}\right),\left[H_{\mathrm{int}, {\rm I}}\left(\tau^{\prime}\right), h_{\mathbf{q}}^s\right]\right] \nonumber\\
         & +\cdots.
	\end{align}
With this formula we expand the Heisenberg picture operators $h_{\mathbf{q},\rm H}^s(\tau)$ into a polynomial consisting of interaction picture operators. Then, we categorize the terms according to the power of the operators in this polynomial.
\begin{equation}
		h_{\mathbf{q},\rm H}^s(\tau)\equiv h_\mathbf{q}^{s,(1)}+h_\mathbf{q}^{s,(2)}+h_\mathbf{q}^{s,(3)}+\cdots,
\end{equation}
can be done here with respect to the number of interaction picture operators in each term, for example, $\mi \int_{\tau_i}^\tau \dif \tau^{\prime}\left[H^{(4)}_{ {\rm I}}\left(\tau^{\prime}\right), h_{\mathbf{q}}^s(\tau)\right]$ includes terms like \(h\delta\phi^2\), which classifies it as a third-order term.
Then, using the unitarity of the evolution operator, $F^\dagger hh F=(F^\dagger hF)(F^\dagger h F)$, the expectation value of $hh$ can also be expand as
\begin{align}
		&\sum_s\left\langle \Omega\left| h_{\mathbf{q},\rm H}^s(\tau) h_{\mathbf{q}^\prime,\rm H}^s(\tau)\right| \Omega\right\rangle \nonumber\\
&=\lim _{\tau_i \rightarrow-\infty(1-\mi \epsilon)}\sum_s \left(\left\langle  h_\mathbf{q}^{s,(1)}h_{\mathbf{q}^\prime}^{s,(1)}\right\rangle+\left\langle  h_\mathbf{q}^{s,(2)}h_{\mathbf{q}^\prime}^{s,(2)}\right\rangle\right. \nonumber\\
 &\left.+\left\langle  h_\mathbf{q}^{s,(1)}h_{\mathbf{q}^\prime}^{s,(3)}+h_\mathbf{q}^{s,(3)}h_{\mathbf{q}^\prime}^{s,(1)}\right\rangle
 +\left\langle  h_\mathbf{q}^{s,(3)}h_{\mathbf{q}^\prime}^{s,(3)}\right\rangle
 +\cdots\right)  \nonumber\\
		&\equiv \left(P_{11}+P_{22}+P_{33}+P_{13}+\cdots\right) (2 \pp)^3 \delta\left(\mathbf{q}+\mathbf{q}^\prime\right),
	\end{align}
where $\left\langle  \right\rangle$ represents expectation values with respect to state $| 0\rangle$.
It is straight forward to compute the tree level power 
spectrum $P_{11}$ and the result is
\begin{equation}
	P_{11}(q)=2\left|v_q(0)\right|^2=\frac{4 H^2}{M_{\mathrm{p}}^2 q^3}.
\end{equation}
\emph{Second order terms}.
Here we present a quick review of the one-loop calculations with the commutator language. Before diving into the lengthy detail, we need to clarify the parameterization we are going to use. In realistic models proposed in ~\cite{Cai:2018tuh,Cai:2019jah,Cai:2019bmk,Zhou:2020kkf,Peng:2021zon,Cai:2021wzd,Inomata:2022ydj}, modes around some certain momentum can be exponentially amplified. Thus inspired us to apply the following ansatz, which a similar form has been used in \cite{Ota:2022hvh}.
We consider the situation that in the momentum space, from a specific $p$-mode $p_*$ to $p_*+d$, the mode functions of $\delta\phi$ are enhanced. In that momentum interval the mode functions are replaced by $u_p(\tau)=\Xi(\tau) u_p^{G}(\tau)$, where
\begin{align}
	\Xi(\tau)=\left\{
	\begin{aligned}
		&0, & \left(\tau<\tau_i\right) \\ &\left(\frac{\tau_i}{\tau}\right)^\mu, & \left(\tau_i\le\tau\le\tau_f\right) \\ &\left(\frac{\tau_i}{\tau_f}\right)^\mu, & \left(\tau_f<\tau\right)
	\end{aligned}\right.
\end{align}
Here the parameter $\mu$ characterizes the Multiples of mode amplification, $\tau_i$ and $\tau_f$ corresponds to the initial and finish of the amplification.
We introduce two Green’s functions $G^G_q$ and $G_q$ to 
simplify the expressions later in this paper
	\begin{align}
		\left[h_{\mathbf{p}}^s\left(\tau^{\prime}\right), h_{\mathbf{q}}^{s_1}(\tau)\right]^\prime&=\left(v_q\left(\tau^{\prime}\right)v_q^*(\tau)-v_q^*\left(\tau^{\prime}\right) v_q(\tau)\right)\delta^{ss_1} \nonumber\\
		&\equiv \frac{4}{M_{\rm p}^2} \frac{\mi}{a(\tau^\prime)^2}G_q^G\left(\tau ; \tau^{\prime}\right)\delta^{ss_1},
	\end{align}
and 
	\begin{align}
		\left[\delta\phi_{\mathbf{p}}\left(\tau^{\prime}\right), \delta\phi_{\mathbf{q}}(\tau)\right]^\prime&=\left(u_q\left(\tau^{\prime}\right)u_q^{*}(\tau)-u_q^{*}\left(\tau^{\prime}\right) u_q(\tau)\right) \nonumber\\
		&\equiv  \frac{\mi}{a(\tau^\prime)^2}G_q\left(\tau ; \tau^{\prime}\right) ,
	\end{align}
where we use prime outside the commutator to denote that the $(2 \pp)^3\delta\left(\mathbf{q}+\mathbf{p}\right)$ part is dropped.

Working out all those commutators, we arrive at
\begin{equation}
	h_\mathbf{q}^{s,(2)}=\frac{2}{M_{\rm p}^2}\int_{\tau_i}^\tau \dif \tau^{\prime} G_q^{G}\left(\tau ; \tau^{\prime}\right) S_{\mathbf{q}}^s\left[\delta \phi\left(\tau^{\prime}\right)\right],
\end{equation}
where $S_{\mathbf{q}}^s$ represents the source term and it’s defined by
\begin{equation}
	S_{\mathbf{q}}^s\left[\delta \phi\left(\tau^{\prime}\right)\right]\equiv\int \frac{\dif ^3 p}{(2 \pp)^3}e_{i j}^{s*}(\hat{q})\mathbf{p}_i \mathbf{p}_j \delta\phi_{\mathbf{p}} \delta\phi_{\mathbf{q}-\mathbf{p}},
\end{equation}
where $e_{i j}^{s*}(\hat{q})$ is the polarization tensor. From $h_\mathbf{q}^{s,(2)}$, one can obtain the $P_{22}$ which is exactly the power spectrum of the so called scalar-induced GWs during inflation. However, from the perspective of Feynman diagrams, $P_{22}$ merely corresponds to a part of Fig. \ref{fey1}. 
\begin{figure}[!htb]
\centering
    \tikzset{every picture/.style={line width=0.75pt}} 
    \begin{tikzpicture}[x=0.75pt,y=0.75pt,yscale=-1,xscale=1]

\draw    (227.99,149.57) .. controls (229.64,147.88) and (231.3,147.86) .. (232.99,149.51) .. controls (234.68,151.15) and (236.34,151.13) .. (237.99,149.44) .. controls (239.64,147.75) and (241.3,147.73) .. (242.99,149.37) .. controls (244.68,151.02) and (246.34,151) .. (247.99,149.31) .. controls (249.64,147.62) and (251.3,147.6) .. (252.99,149.24) .. controls (254.68,150.89) and (256.34,150.87) .. (257.99,149.18) .. controls (259.64,147.49) and (261.3,147.47) .. (262.99,149.11) .. controls (264.68,150.75) and (266.34,150.73) .. (267.99,149.04) .. controls (269.64,147.35) and (271.3,147.33) .. (272.99,148.98) -- (276.08,148.94) -- (276.08,148.94) ;
\draw [shift={(276.08,148.94)}, rotate = 359.25] [color={rgb, 255:red, 0; green, 0; blue, 0 }  ][fill={rgb, 255:red, 0; green, 0; blue, 0 }  ][line width=0.75]      (0, 0) circle [x radius= 2.01, y radius= 2.01]   ;
\draw    (326.08,148.94) .. controls (327.71,147.25) and (329.38,147.22) .. (331.07,148.86) .. controls (332.76,150.5) and (334.43,150.47) .. (336.07,148.78) .. controls (337.71,147.09) and (339.38,147.06) .. (341.07,148.7) .. controls (342.76,150.34) and (344.43,150.31) .. (346.07,148.62) .. controls (347.71,146.93) and (349.38,146.9) .. (351.07,148.54) .. controls (352.76,150.18) and (354.43,150.15) .. (356.07,148.46) .. controls (357.71,146.77) and (359.38,146.74) .. (361.07,148.37) .. controls (362.76,150.01) and (364.43,149.98) .. (366.07,148.29) -- (366.95,148.28) -- (366.95,148.28) ;
\draw [shift={(326.08,148.94)}, rotate = 359.08] [color={rgb, 255:red, 0; green, 0; blue, 0 }  ][fill={rgb, 255:red, 0; green, 0; blue, 0 }  ][line width=0.75]      (0, 0) circle [x radius= 2.01, y radius= 2.01]   ;
\draw   (276.08,148.94) .. controls (276.08,135.13) and (287.27,123.94) .. (301.08,123.94) .. controls (314.88,123.94) and (326.08,135.13) .. (326.08,148.94) .. controls (326.08,162.75) and (314.88,173.94) .. (301.08,173.94) .. controls (287.27,173.94) and (276.08,162.75) .. (276.08,148.94) -- cycle ;
\end{tikzpicture}
    \caption{Diagram corresponds to $P_{22}$ and $P_{13a}$\label{fey1}}
\end{figure}\\
Meanwhile, As emphasized in \cite{Ota:2022hvh,Ota:2022xni}, $P_{13}$ cannot be neglected since it is also of one-loop order, just as $P_{22}$ does. To compute this cross-correlation term, $h_\mathbf{q}^{s,(3)}$ must be considered, and the results are given by
\begin{subequations}
	\begin{align}
		&h_\mathbf{q}^{s,(3)}=h_\mathbf{q}^{s,(3a)}+h_\mathbf{q}^{s,(3b)},&\\
        & h_\mathbf{q}^{s,(3a)} =\frac{2}{M_{\rm p}^2}\int_{\tau_i}^\tau \dif \tau^{\prime}\int_{\tau_i}^{\tau^\prime}\dif \tau^{\prime\prime} G_q^G\left(\tau ; \tau^{\prime}\right) \nonumber\\
		&\times   \int \frac{\dif^3 p \dif^3 k}{(2 \pp)^6}e_{i j}^{s*}(\hat{q})\mathbf{p}_i \mathbf{p}_j  G_p\left(\tau^\prime ; \tau^{\prime\prime}\right) \sum_{s_1} h_\mathbf{k}^{s_1}(\tau^{\prime\prime}) \nonumber\\
		&\times e_{mn}^{s_1}(\hat{k})\mathbf{p}_m \mathbf{p}_n  \{\delta\phi_{\mathbf{p}-\mathbf{k}}(\tau^{\prime\prime}) \delta\phi_{\mathbf{q}-\mathbf{p}}(\tau^{\prime})\}, \\
        &h_\mathbf{q}^{s,(3b)}=-\frac{2}{M_{\rm p}^2}\int_{\tau_i}^\tau \dif \tau^{\prime}G_q^G\left(\tau ; \tau^{\prime}\right)\int \frac{\dif^3 p \dif^3 k}{(2 \pp)^6} \sum_{s_2}e_{i k}^{s*}(\hat{q}) \nonumber\\
		&\times e_{k j}^{s_2}(\hat{p}) \mathbf{k}_i(\mathbf{q}-\mathbf{k})_j \delta\phi_\mathbf{k}(\tau^{\prime})\delta\phi_{\mathbf{q}-\mathbf{p}-\mathbf{k}}(\tau^{\prime})h_\mathbf{p}^{s_2}(\tau^{\prime}),
	\end{align}
\end{subequations}
where the expression is divided into two parts corresponding to $H^{(3)}$ and $H^{(4)}$, respectively.

\begin{figure}[!htb]
    \centering
\begin{tikzpicture}[x=0.75pt,y=0.75pt,yscale=-1,xscale=1]

\draw    (248.39,169.63) .. controls (250.04,167.96) and (251.71,167.95) .. (253.39,169.6) .. controls (255.06,171.26) and (256.73,171.25) .. (258.39,169.58) .. controls (260.05,167.91) and (261.72,167.9) .. (263.39,169.55) .. controls (265.06,171.2) and (266.73,171.19) .. (268.39,169.52) .. controls (270.05,167.85) and (271.72,167.84) .. (273.39,169.49) .. controls (275.06,171.14) and (276.73,171.13) .. (278.39,169.46) .. controls (280.05,167.79) and (281.72,167.78) .. (283.39,169.43) .. controls (285.06,171.08) and (286.73,171.07) .. (288.39,169.4) .. controls (290.05,167.73) and (291.72,167.72) .. (293.39,169.37) .. controls (295.06,171.02) and (296.73,171.01) .. (298.39,169.34) .. controls (300.05,167.67) and (301.72,167.66) .. (303.39,169.31) .. controls (305.06,170.97) and (306.73,170.96) .. (308.39,169.29) .. controls (310.05,167.62) and (311.72,167.61) .. (313.39,169.26) .. controls (315.06,170.91) and (316.73,170.9) .. (318.39,169.23) .. controls (320.05,167.56) and (321.72,167.55) .. (323.39,169.2) .. controls (325.06,170.85) and (326.73,170.84) .. (328.39,169.17) .. controls (330.05,167.5) and (331.72,167.49) .. (333.39,169.14) .. controls (335.06,170.79) and (336.73,170.78) .. (338.39,169.11) .. controls (340.05,167.44) and (341.72,167.43) .. (343.39,169.08) .. controls (345.06,170.73) and (346.73,170.72) .. (348.39,169.05) -- (350.18,169.04) -- (350.18,169.04) ;
\draw [shift={(299.29,169.34)}, rotate = 359.67] [color={rgb, 255:red, 0; green, 0; blue, 0 }  ][fill={rgb, 255:red, 0; green, 0; blue, 0 }  ][line width=0.75]      (0, 0) circle [x radius= 2.01, y radius= 2.01]   ;
\draw   (276.44,146.32) .. controls (276.44,133.59) and (286.76,123.27) .. (299.49,123.27) .. controls (312.22,123.27) and (322.54,133.59) .. (322.54,146.32) .. controls (322.54,159.05) and (312.22,169.37) .. (299.49,169.37) .. controls (286.76,169.37) and (276.44,159.05) .. (276.44,146.32) -- cycle ;
\end{tikzpicture}
    \caption{Diagram corresponds to $P_{13b}$\label{fey2}}
    \label{fig:enter-label}
\end{figure}
This contribution to the one-loop power spectrum consists of two terms, $P_{13a}$ and $P_{13b}$. Specifically, $P_{13a}$ corresponds to the other part of Fig.\ref{fey1} beyond $P_{22}$, while $P_{13b}$ corresponds to Fig.\ref{fey2}. 

Using these results, we obtain integral expressions for the one-loop power spectrum through several contractions. We first address the momentum integrals, which involve a triple integral over the entire momentum space.

\begin{figure}[!htb] 
	\centering 
	\includegraphics[width=0.48\textwidth]{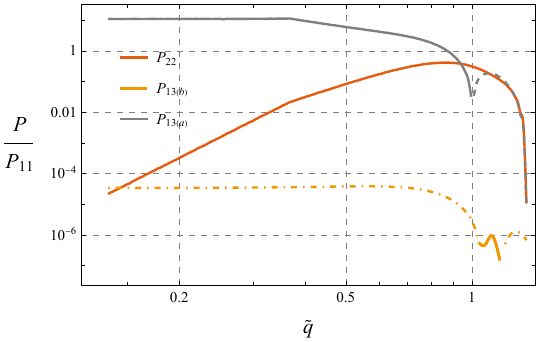} 
	\label{fig:11} 
	\caption{ One loop order tensor power spectrum including $P_{22}$ and $P_{13}$, where they have been normalized by dividing $P_{11}$. In this figure, we denote positive values with solid lines  and negative values with dashed lines. Initial and final times of the growth are fixed to $x_i = p_*\tau_i = -30$ and $x_f = p_*\tau_f = -3$, respectively, while we take $\mu=3.2$. Also, we also used the normalized momentum $\tilde{q}=\frac{q}{p_*}$.\label{one loop}}
\end{figure}

Following \cite{Ota:2022xni}, we expect that the dominant contributions arise from the enhanced modes, as other  modes are in  ground state. Therefore, contributions from those ground state modes can be approximately neglected thus we only integrate over momentum interval $p_*$ to $p_*+d$. By applying appropriate approximations to the integration region, we analytically perform these momentum integrals. The details of our approximations are provided in the Appendix \ref{appendixA}. Finishing momentum integrals yields the integrand of the time integral, which we can evaluate numerically. We fix the initial and final time of the growth while varying the index $\mu$ to adjust the magnification. 
In order to figure out the momentum dependence of one-loop order power spectrum, we have chosen a specific magnification index $\mu$ and calculated $P_{13(a(b))}$, which corresponds to correlations between $h_\mathbf{q}^{s,1}$ and $h_\mathbf{q}^{s,(3a(b))}$ respectively as shown in Fig. \ref{one loop}.

As illustrated by \cite{Ota:2022xni}, there are several noteworthy aspects about these results. We are already familiar with the $\mathcal{O}(q^3)$ infrared (IR) behavior of the $P_{22}$ terms, which arises naturally from the causal production of GWs. However, the $P_{13}$ terms exhibit scale invariance for IR modes. To explore the IR behavior, we consider the limit where \( |q\tau| \ll 1 \). In this regime, Green function can be approximated as
\begin{equation}
	G^G_q\left(0; \tau^{\prime}\right) = -\frac{4 H^2 a^{\prime 2} \tau^{\prime 3}}{3 M_{\mathrm{pl}}^2} \left[1 + \mathcal{O}\left(\tau^{\prime 2} q^2\right)\right].
\end{equation}

Further analysis reveals that for \( q \ll p_* \) and \( |q\tau| \ll 1 \), \( |q\tau'| \ll 1 \), the expression \( S_{\mathbf{q}}^s[\delta\phi] \) can be written as \( e^{i j, s, *}(\hat{q}) \Sigma_{i j}[\delta\phi] \), where \( \Sigma_{i j}[\delta\phi] \) is almost scale-invariant. Consequently, we find that \(\displaystyle \frac{P_{22}}{P_{11}} = \mathcal{O}\left(\tilde{q}^3\right) \)  on large scales. Notably, there is an "additional" term of the form \( v_\mathbf{q}(\tau)v^*_\mathbf{q}(\tau^\prime) \) in the expression for \( P_{13} \), which exactly cancels out the \( q \) dependence introduced when dividing by \( P_{11} \), resulting in a  \( \mathcal{O}\left(\tilde{q}^0\right) \) IR behavior.

It is noteworthy that $P_{13}\approx P_{13(a)}$ , particularly in the infrared (IR) limits. This observation indicates that the traditional reliance on the number of loops for quantifying orders of magnitude may be misleading. A more precise approach involves considering the exponent of the ratio ${H}/{M_{\rm p}}$ alongside the enhancement factor $\Xi$. For instance, $P_{13(a)}$ is of order $\left({H}/{M_{\rm p}}\right)^2 \Xi^4$ which is of same order as $P_{22}$, we can mark their order by $(2,4)$, while $P_{13(b)}$ is only of order $(2,2)$. For the large-scale spectrum, it is straightforward to verify that this approximation aligns well with the numerical results, effectively capturing the underlying behavior.

\begin{figure}[!htb]
	\centering
	\subfigure[]{ \includegraphics[width=.45\textwidth]{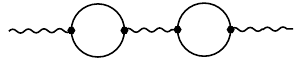}\label{fig:Fm-1}}
	\subfigure[]{ \includegraphics[width=.25\textwidth]{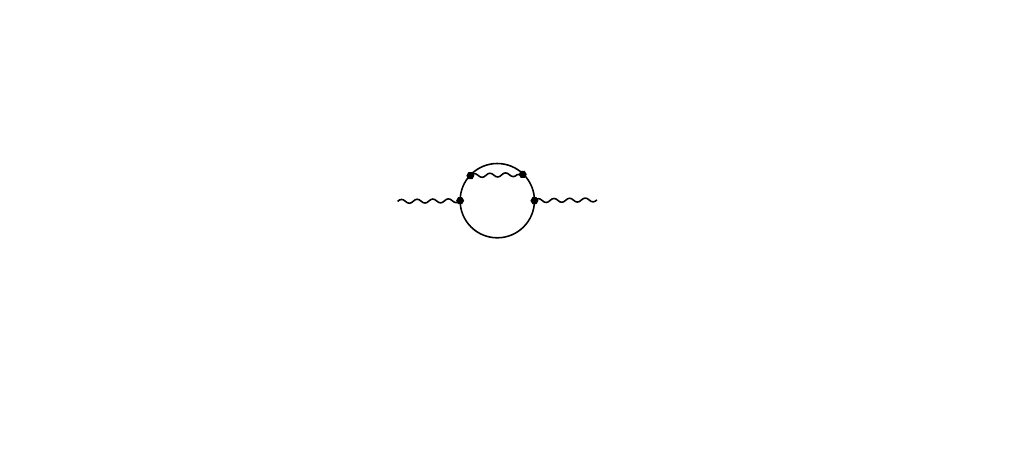}\label{fig:Fm-2}}
	\subfigure[]{ \includegraphics[width=.25\textwidth]{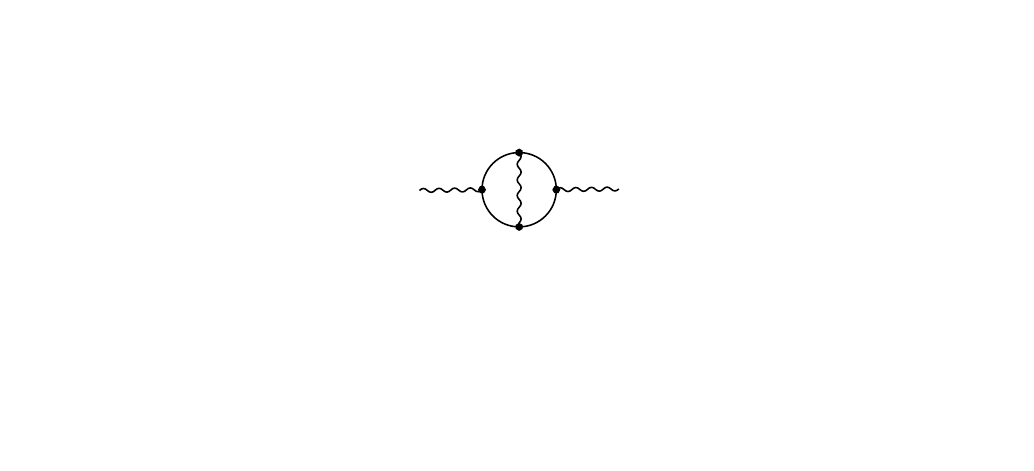}\label{fig:Fm-3}}
	\caption{Three diagrams correspond to $P_{33aa}$}
	\label{two loop diagram}
\end{figure}

\emph{Third order spectrum}.
The counter-correlation terms exhibit an interesting characteristic: Fig. \ref{one loop} clearly demonstrates that the large-scale values of $P_{13}$ is significantly larger, even exceeding the peak value of $P_{22}$, which contrasts sharply with the corresponding terms in the RD era calculated in \cite{Chen:2022dah}. The significant magnitude of $\left|\left\langle h_{\mathbf{k}}^{(1)} h_{\mathbf{-k}}^{(3)}\right\rangle\right|$ indicates that maybe the $\left\langle h_{\mathbf{k}}^{(3)} h_{\mathbf{-k}}^{(3)}\right\rangle$ terms themselves can not be safely neglected. It is straight forward to divide these terms into three parts: $P_{33aa}$, $P_{33ab}$, and $P_{33bb}$, which are of order $(4,8)$, $(4,6)$, and $(4,4)$, respectively. The order of these terms indicates that $P_{33ab}$, and $P_{33bb}$ can be safely neglected comparing to $P_{33aa}$, thus from now on we will focus on $P_{33aa}$. Upon handling it's contractions, we identify three types of momentum structures. The diagrammatical expression of $P_{33}\approx P_{33aa}$ are shown in Fig. \ref{two loop diagram}. 

 In order to have a clearer comparison of $P_{13}$ and $P_{33}$, we take the IR limit and consider their behavior with respect to various $\mu$. 
 It is easy to identify that only Fig.\ \ref{fig:Fm-1} has a \(\mathcal{O}\left(\tilde{q}^0\right)\) momentum dependence. The reason is that the momentum of the inner tensor propagator is fixed to $q$ by the conservation of 3-momentum, which leaves a $ v_q(\tau^{\prime})v^*_{q}(\tau^{\prime\prime}) $ term in the expressions. For the other two diagrams, the momentum of inner $h$ is integrated over the entire momentum space so they will obtain a $\tilde{q}^3$ depression for large scales after being normalized by $P_{11}$.

As a result, we consider the third order contribution to the GW power spectrum from the Fig.\ \ref{fig:Fm-1}. The expression can be written as
	\begin{align}
		&\left\langle h_{\mathbf{q}}^{r} h_{\mathbf{q}^\prime}^{s}\right\rangle=(2 \pp)^3 \delta\left(\mathbf{q}+\mathbf{q}^\prime \right)\frac{4}{M_{\rm p}^4}\int_{\tau_i}^\tau \dif \tau^{\prime}\int_{\tau_i}^{\tau^\prime} \dif \tau^{\prime\prime}\nonumber\\
		&\times G_{q}^G\left(\tau ; \tau^{\prime}\right)\int_{\tau_i}^{\tau_1} \dif \tau_1^{\prime}\int_{\tau_i}^{\tau_1^\prime}\dif {\tau_1}^{\prime\prime} G_{q}^G\left(\tau_1 ; \tau_1^{\prime}\right)v_q(\tau^{\prime\prime})v^*_{q}(\tau_1^{\prime\prime})\nonumber\\
		&\times\sum_{s_1}\int \frac{\dif^3 p}{(2 \pp)^3}2 e_{i j}^{r}(\hat{q})\mathbf{p}_i \mathbf{p}_j e_{mn}^{s_1^*}(\hat{q})\mathbf{p}_m \mathbf{p}_n G_q\left(\tau^\prime ; \tau^{\prime\prime}\right)\nonumber \\
		&\times \Re(u_{|\mathbf{q}-\mathbf{p}|}(\tau_1^{\prime\prime})u_{|\mathbf{q}-\mathbf{p}|}^*(\tau_1^{\prime}))\int \frac{\dif^3 k}{(2 \pp)^3}e_{\alpha\beta}^{s*}(\hat{q})\mathbf{k}_{\alpha}\mathbf{k}_{\beta}e_{kl}^{s_1}(\hat{q})\nonumber\\
		&\times \mathbf{k}_{k}\mathbf{k}_{l}G_q\left(\tau_1^\prime ; \tau_1^{\prime\prime}\right)2\Re \left[u_{|\mathbf{q}-\mathbf{k}|}(\tau_1^{\prime\prime})u_{|\mathbf{q}-\mathbf{k}|}^*(\tau_1^{\prime})\right],
	\end{align}
Following the method used to calculate the one-loop contributions, the results are straightforward.
\begin{figure}[htbp] 
	\centering 
	\includegraphics[width=0.48\textwidth]{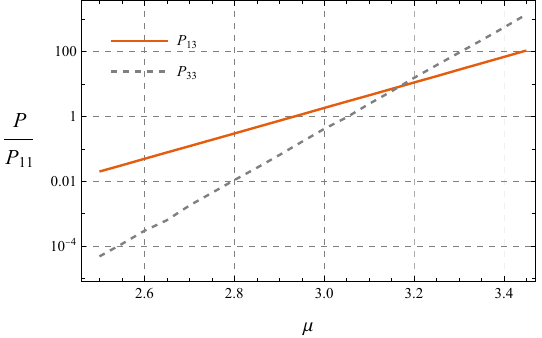} 
	\caption{Infrared values of $P_{13}$ and $P_{33}$, where the spectrum have also been normalized. }
		\label{loop value}
\end{figure}

In Fig. \ref{loop value}, we compare the IR values of $P_{13}$ and $P_{33}$ for various $\mu$, where we can clearly find that when $P_{13}$ are strictly smaller than linear terms, two-loop contributions are even much smaller. However, as we gradually enhance the scalar perturbations, $P_{33}$ is more intensely enhanced and finally exceed $P_{13}$. There's no doubt that this result agree well with the fact that, as the scalar modes become larger$P_{33}$ grows as  $\Xi^4$ order and  $P_{13}$ grows as $\Xi^2$ order.

\emph{Conclusion and discussion}.
As we have mentioned above, the IR value of $P_{13}$ is always larger than $P_{22}$, so if $\delta\phi$ are enhanced large enough to make the third order corrections even larger, we have to admit that $h_\mathbf{q}^{s,(3)}$ can't be neglected in those circumstances, which means $P_{22}$ can't be consider as a good approximation of the physical observable. More numerical experiments also showed that $P_{22}$ can't even exceed the linear order spectrum with the request that the GWs are perturbatively induced. 
While we have proved that the third order terms are larger during inflation through direct calculations, one may wonder why this didn't happen in the RD era. The key point is that during inflation, the Green function of scalar modes are of $\Xi^2$ order, while when we consider the scalar-induced GWs in the RD era, only the initial condition not the Green functions are enhanced. 

In more realistic models, this kind of enhancement corresponds to changing of the linear order equations of motions (EoMs), for example, in the widely considered resonance models \cite{Cai:2018tuh,Cai:2019jah,Cai:2019bmk,Zhou:2020kkf,Peng:2021zon,Inomata:2022ydj}, a tiny oscillating structure is added to the inflaton potential which contributes a resonance term in the EoM of inflaton perturbations. 
On the other hand, in our parameterization, during the period of time that the oscillating term is active, the mode function of the scalar perturbations will be exponentially enhanced, which is similar to those resonance models. However, we are not sure whether the results will change for more realistic models, we will do further theoretical and numerical examinations in our future works.

Actually, there's another perspective of our results, if we consider the higher order corrections with the EoM method, corresponding three order tensor modes can be expressed as
\begin{align}
    &{h_{\mathbf{q}}^s}^{(3)}=\frac{2}{M_{\rm p}^2}\int^\tau \dif\tau' G^{\rm BD}_k(\tau,\tau') \nonumber \\
    &\times \int \frac{\dif^3 p}{(2 \pp)^3}e_{i j}^{s*}(\hat{q})\mathbf{p}_i \mathbf{p}_j\delta\phi_{\mathbf{p}}^{(2)}\delta \phi_{\mathbf{q}-\mathbf{p}},
\end{align}
where
\begin{align}
    &\delta\phi_{\mathbf{q}}^{(2)}=\int^\tau \dif\tau' G_{k}(\tau,\tau') \nonumber \\
    &\times \int \frac{\dif^3 p}{(2 \pp)^3}\sum_s e_{i j}^{s}(\hat{p})\mathbf{q}_i \mathbf{q}_j h_{\mathbf{p}}^{s} \delta\phi_{\mathbf{q}-\mathbf{p}}.
\end{align}
We can surprisingly find that the order of $\left\langle\delta \phi_{-\mathbf{p}}^{(2)} \delta\phi_{\mathbf{p}}^{(2)}\right\rangle$ is the same as $\left\langle h_{-\mathbf{p}}^{(2)} h_{\mathbf{p}}^{(2)}\right\rangle$, so if the power spectrum of the induced GWs are much larger than  $P_{11}$, the scalar perturbation itself is also nonperturbative. It is also instructive here to compare our calculations with the results presented in \cite{Inomata:2022yte}, where the self-interaction of the scalar field perturbations are considered. These interactions also contributes considerable loop corrections when the scalar modes are enhanced large enough and thus give perturbative constraints on the scalar magnification. According to their results, tree-level scalar perturbations must be amplified by approximately $10^7$ times before loop corrections dominate over tree-level results, leaving room for exploring scalar - tensor interactions in our framework. Although both of them considered higher-order scalar two-point correlation functions, we argue that the effect we are discussing is more general because it is an intrinsic property of the GW-production, while the calculations of \cite{Inomata:2022yte} rely on the details of the resonance models. In this letter, we proved directly that nonperturbative effects come from exactly the same mechanism. It is also worth noticing that recent studies on small-scale nonlinear processes during inflation have advanced to the level of lattice simulations \cite{Caravano:2024tlp}, however, future works are still needed to perform lattice simulations of the tensor modes.

More Intuitively, the Fig.\ \ref{fig:Fm-2} of $\left\langle h_{-{\mathbf{p}}}^{(3)} h_{\mathbf{p}}^{(3)}\right\rangle$ is directly related to the self-correlations of the second order $\phi$; its values are shown below, where we can also see the two-loop spectrum exceed the one-loop spectrum as the scalar perturbations become larger.  

\begin{figure}[htbp]
	\centering
	\includegraphics[width=0.48\textwidth]{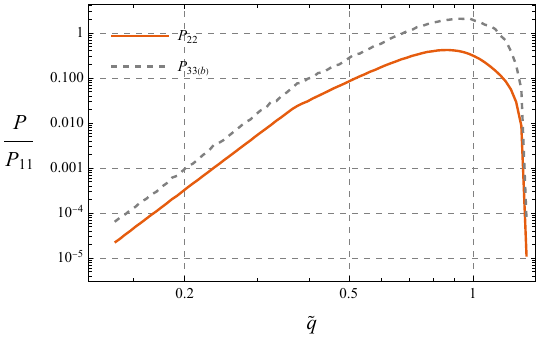}
	\label{fig:44}
	\caption{Power spectrum corresponds to Fig.\ \ref{fig:Fm-2} of $P_{33}$ and $P_{22}$, where the spectrum and momentum are also normalized.}
\end{figure}

It should be mentioned here that we haven't considered all two-loop diagrams, which also include cross-correlation terms like $\left\langle h_{-\mathbf{p}}^{(2)} h_{\mathbf{p}}^{(4)}\right\rangle$ and $\left\langle h_{-\mathbf{p}}^{(1)} h_{\mathbf{p}}^{(5)}\right\rangle$. However, using the order parameter we mentioned above, we find that they are of the same order. Besides, upon examining Feynman diagrams with increasing numbers of loops, we discovered that they incorporate more enhancement factors. Irrespective of the number of loops, there consistently exist terms of order $(2n, 4n)$, where $n$ is the number of loops. Thus they may grow even faster than two loop terms. This indicates that in such a case, considering the perturbative expansion of two-loop diagrams or even higher order diagrams will not yield the correct result, and the perturbative method has completely failed here.These facts indicate that perturbative calculations for observable scalar induced GWs during inflation era may be impossible.

\emph{Acknowledgements. } This work is supported in part by the National Key Research and Development Program of China Grant No. 2020YFC2201501, in part by the National Natural Science Foundation of China under Grant No. 12075297 and No. 12235019.

\appendix
\section{Analytical calculation of momentum integral in one-loop contribution}\label{appendixA}.
\subsection{Linear order}
The Fourier modes of the scalar field perturbation and tensor perturbation are expressed as
\begin{align}
	\delta\phi(\tau, \mathbf{x}) & =\int \frac{\dif^3 k}{(2 \pp)^3} \me^{\mi \mathbf{k} \cdot \mathbf{x}} \delta\phi_{\mathbf{k}}(\tau) \\
	h_{i j}(\tau, \mathbf{x}) & =\int \frac{\dif^3 k}{(2 \pp)^3} \me^{\mi \mathbf{k} \cdot \mathbf{x}} \sum_{s= \pm 2} e_{i j}^s(\mathbf{k}) h_{\mathbf{k}}^s(\tau).
\end{align}
Taken only quadratic Hamiltonian in to consideration, we can first solve the time evolution of $\phi_{\mathbf{k}}$ and $h_{\mathbf{k}}^s$ in the interaction picture. According to standard quantization procedures, these perturbation terms can be expand in to creation and annihilation operators as
\begin{align}
	\delta\phi_{\mathbf{q}}(\tau) & =u_q(\tau) a_{\mathbf{q}}+u_q^*(\tau) a_{-\mathbf{q}}^{\dagger}, \\
	h_{\mathbf{q}}^s(\tau) & =v_q(\tau) b_{\mathbf{q}}^s+v_q^*(\tau) b_{-\mathbf{q}}^{s \dagger},
\end{align}
where the $u_q(\tau)$ and $v_q(\tau)$ are mode functions, and the annihilation and creation operators satisfy
\begin{align}
	&a_{\mathbf{q}}|0\rangle=b_{\mathbf{q}}^s|0\rangle=0\\
	&{\left[a_{\mathbf{q}}, a_{-{\mathbf{q}^\prime}}^{\dagger}\right] } =(2 \pp)^3 \delta(\mathbf{q}+{\mathbf{q}^\prime}) \\
	&{\left[b_{\mathbf{q}}^s, b_{-{\mathbf{q}^\prime}}^{r \dagger}\right] }  =(2 \pp)^3 \delta^{sr} \delta(\mathbf{q}+{\mathbf{q}^\prime}).
\end{align}
For ground states that the scalar perturbations are not amplified, the mode functions are given by
\begin{align}
	u_k^{\mathrm{G}}(\tau) & =\frac{H}{\sqrt{2 k^3}}(1+\mi k \tau) \me^{-\mi k \tau}, \\
	v_k^{\mathrm{G}}(\tau) & =\frac{2 H}{M_{\mathrm{p}} \sqrt{2 k^3}}(1+\mi k \tau) \me^{-\mi k \tau}.
\end{align}
Hence we can get the linear order power spectrum through direct calculations
\begin{equation}
	P_{11}(q)=2\left|v_q(0)\right|^2=\frac{4 H^2}{M_{\mathrm{p}}^2 q^3}.
\end{equation}

\subsection{Interaction picture}

The evolution operator in the interaction picture are defined as
\begin{equation}
	\begin{aligned}
		F\left(\tau ; \tau_i\right)&=U_0^{-1}\left(\tau, \tau_0\right) U\left(\tau, \tau_0\right)
	\end{aligned}
\end{equation}
where the $U_0\left(\tau, \tau_0\right)$ and $U\left(\tau, \tau_0\right)$ denote evolution operator correspond to free and full hamiltonian respectively. We can get directly from definition that this operator satisfy
\begin{equation}
	\partial_\tau\left[U_0^{-1}\left(\tau, \tau_0\right) U\left(\tau, \tau_0\right)\right]=-\mi U_0^{-1}\left(\tau, \tau_0\right) U\left(\tau, \tau_0\right) H_{\mathrm{int}},
\end{equation}
which can be solved iteratively
	\begin{align}
		F\left(\tau ; \tau_i\right)&=\mathbb{T} \exp \left(- \mi \int_{\tau_0}^\tau \dif \tau^{\prime \prime} H_{\mathrm{int}, {\rm I}}\left(\tau^{\prime \prime}\right)\right)\nonumber\\
		&=1-\mi \lambda \int_{\tau_i}^\tau \dif \tau^{\prime} H_{\mathrm{int}, {\rm I}}\left(\tau^{\prime}\right) \nonumber\\
		&-\lambda^2 \int_{\tau_i}^\tau \dif \tau^{\prime} \int_{\tau_i}^{\tau^{\prime}} \dif \tau^{\prime \prime} H_{\mathrm{int}, {\rm I}}\left(\tau^{\prime}\right) H_{\mathrm{int}, {\rm I}}\left(\tau^{\prime \prime}\right)\nonumber\\
        &+\mathcal{O}\left(\lambda^3\right),
	\end{align}
note that the order counting parameter $\lambda$ is shown explicitly.

\subsection{Momentum integrals}

As mentioned in the previous chapter, only enhanced k-modes are included in our momentum integrals. Here, I'd like to used two examples to illustrate our method. First, the momentum part of $P_{22}$ can be written as
\begin{equation}
	\int \frac{\dif^3 p}{(2 \pp)^3}e^{s*}(q,p)e^{s'}(q,p) u_{p}(\tau^{\prime})f(|\mathbf{q}-\mathbf{p}|,p),
\end{equation}
where $e^{s*}(q,p)=e_{i j}^{s*}(\hat{q})\mathbf{p}_i \mathbf{p}_j $ and
\begin{equation}
    f(|\mathbf{q}-\mathbf{p}|,p)=u_{{p}}^*(\tau^{\prime\prime})u_{|\mathbf{q}-\mathbf{p}|}(\tau^{\prime})u_{|\mathbf{q}-\mathbf{p}|}^*(\tau^{\prime\prime}).
\end{equation}
At first glance, a reasonable approximation for narrow spectrum is 
\begin{equation}
	f\left(|\mathbf{q}-\mathbf{p}|, p\right)=\delta\left[\ln \left(\frac{|\mathbf{q}-\mathbf{p}|}{p_*}\right)\right] \delta\left[\ln \left(\frac{p}{p_*}\right)\right] f\left(p_*, p_*\right),
\end{equation}
with this condition, the momentum integral can be simplified as a angle integral
\begin{equation}
	\int\frac{\dif \theta}{(2 \pp)^3} p_*^8 \pp \frac{\sin^4{\theta}}{q} \delta\left(\theta-\theta_0\right)=\frac{p_*^4 \pp}{q}\left(p_*^2-\frac{q^2}{4}\right)^2,
\end{equation}
this result will finally leads us to a $\mathcal{O}(q^2)$ infrared behaviour which should not happen in local production of GWs. We'd like to consider a more physical setting to reveal the problems hidden behind the $\delta$ approximation and to obtain a more reasonable result. 
We require that only from a specific $p$-mode $p_*$ to $p_*+d$, the mode functions of $\phi$ are enhanced, thus we have two characteristic momentum scale $q$ and $d$. 
We consider the integration region $\Omega$ in the momentum space, it's the overlap part of two spherical shell as shown in Fig. 4.
\begin{equation}
	\Delta\Pi=\int_\Omega\frac{\dif p\dif \theta}{(2 \pp)^3}\pp p^6 \sin^5{\theta}
\end{equation}
For $d\ll q$, the integration region is a ring and the polar angle only vary in a very narrow interval around $\theta_0$, which is similar to the $\delta$ case. However, no matter how small the width $d$ is, there always exist $q<d$, and for these infrared $q$ we must integrate over a shell which means $\theta$ vary from $0$ to $\pp$. It still difficult to obtain the exact value of this integral, so we decided to approximate it from two limits and link them up at $q=d$, finally we adopt the following results
\begin{equation}
	\Delta\Pi=\left\{
	\begin{aligned}
		&\frac{\pp p_*^6 d^2 \sin^4{\theta_0}}{(2 \pp)^3 q}, & \left(q<d\right) \\ &\frac{\pp p_*^6 d }{(2 \pp)^3 }, & \left(q>d\right).
	\end{aligned}\right.
\end{equation}
and now the momentum integral of $P_{22}$ have been completed. In the second example we are going to deal with the momentum part of Fig.\ \ref{fig:Fm-2} of $P_{33aa}$. We first give the full expressions of $\left\langle h_{-\mathbf{p}}^{(3)} h_{\mathbf{p}}^{(3)}\right\rangle^{(b)}$
	\begin{align}
		&\left\langle h_{\mathbf{q}}^{r} h_{\mathbf{q}^\prime}^{s}\right\rangle=(2 \pp)^3 \delta\left(\mathbf{q}+\mathbf{q}^\prime\right)
		\frac{4}{M_{\rm p}^4}\int_{\tau_i}^\tau \dif \tau^{\prime}G_q^G\left(\tau ; \tau^{\prime}\right) \nonumber\\
		&\int_{\tau_i}^{\tau} \dif \tau_1^{\prime} G_q^G\left(\tau ; \tau_1^{\prime}\right)\int \frac{\dif^3 p}{(2 \pp)^3}e_{i j}^{r}(\hat{q})\mathbf{p}_i \mathbf{p}_j e_{mn}^{s^*}(\hat{q})\mathbf{p}_m \mathbf{p}_n \nonumber\\
		&u_{|\mathbf{q}-\mathbf{p}|}(\tau^{\prime})u_{|\mathbf{q}-\mathbf{p}|}^*(\tau_1^{\prime}) U(\tau^{\prime},\tau_1^{\prime}),
	\end{align}
where
	\begin{align}
		&U(\tau^{\prime},\tau_1^{\prime})=4\int_{\tau_i}^{\tau^\prime} \dif \tau^{\prime\prime} \int_{\tau_i}^{\tau_1^\prime}\dif {\tau_1}^{\prime\prime} G_p\left(\tau^\prime ; \tau^{\prime\prime}\right) G_p\left(\tau_1^\prime ; \tau_1^{\prime\prime}\right)\nonumber \\
		&\times \int \frac{\dif^3 k}{(2 \pp)^3} \sum_{s_2}|e_{\alpha\beta}^{s_2}(\hat{k})\mathbf{p}_{\alpha}\mathbf{p}_{\beta}|^2 v_k(\tau^{\prime\prime}) \nonumber\\
        & \times v^*_k(\tau_1^{\prime\prime}) u_{|\mathbf{p}-\mathbf{k}|}(\tau^{\prime\prime})u_{|\mathbf{p}-\mathbf{k}|}^*(\tau_1^{\prime\prime}).
	\end{align}
Noticed that the integral of $p$ is similar to the one-loop diagram, we will focus on the $k$ integral which is independent of $q$ so the $\delta$ approximation can be safely applied, but this time the normalization factor have been chosen to be $d$. After some direct calculations, only a integral of $\theta$ is left
\begin{equation}
	Y=\int_{\frac{\pp}{2}}^\pp \dif \theta \frac{\sin^5{\theta}}{\cos^2{\theta}} (1+\mi k\tau^{\prime\prime}) (1-\mi k\tau_1^{\prime\prime}) \me^{\mi k(\tau_1^{\prime\prime}-\tau^{\prime\prime})},
\end{equation}
where $k=-2p_*\cos{\theta}$. Unfortunately $Y$ goes to infinity as $\theta$ approaches ${\pp}/{2}$, which means we are facing a infrared divergence and we have to regularize it. To do that we introduce a positive parameter $\theta_0$ by hand and integrate it analytically
	\begin{align}
		Y(\theta_0)&=\int_{\frac{\pp}{2}+\theta_0}^\pp \dif \theta \frac{\sin^5{\theta}}{\cos^2{\theta}} (1+\mi k\tau^{\prime\prime}) (1-\mi k\tau_1^{\prime\prime}) \me^{\mi k(\tau_1^{\prime\prime}-\tau^{\prime\prime})} \nonumber \\
		&=-\frac{1}{\theta_0}+Y_0+Y_1 \theta_0+\mathcal{O}(\theta_0^2).
	\end{align}
To do complete the regularization process we just need to subtract the divergent term and take $\theta_0 \rightarrow 0$ so that $Y_0$ is just what we need.

	\bibliographystyle{apsrev4-1}
\bibliography{pertGW}

\end{document}